\documentstyle[epsfig]{prhep97}

\twocolumn 

\makeatletter
\let\chapter\hid@chapter
\makeatother
\begin{document}
\setlength{\unitlength}{1cm}


\authorrunning{A.\,Lindner}
\titlerunning{{\talknumber}: New Results of HEGRA}
 

\def\talknumber{1407} 

\title{{\talknumber}: New Results of the HEGRA Air Shower Detector Complex }
\author{Axel\,Lindner
(axel.lindner@desy.de), for the HEGRA 
Collaboration$^1$}

\institute{University of Hamburg, Germany}

\maketitle

\vspace*{-0.5cm}
\begin{abstract}
Recent highlight results obtained with 
the air shower 
detector complex HEGRA at the Canary island La Palma
are presented.
Observations and searches for 
${\rm \gamma}$ sources above an energy of 
500\,GeV are covered,
where special emphasis is put on the recent 
flaring activities of the Blazar Mkn 501. 
A second main research topic of the experiment is the 
measurement of the energy
spectrum and the elemental composition of the charged cosmic rays.
Data covering the energy region of the "knee" 
around 3\,PeV are presented.
\vspace*{-0.2cm}
\end{abstract}
\addtocounter{footnote}{1}
\vspace*{-0.5cm} 
\section{The Setup}
\begin{figure}
\vspace*{-1.5cm}
\epsfig{file=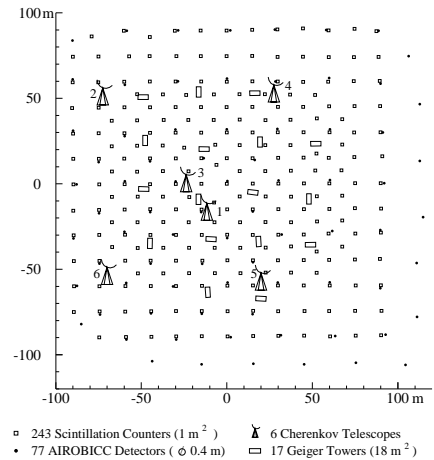,width=5.6cm,angle=0}
\vspace*{-1.5cm}
\caption{\small \em The layout of the HEGRA experiment at La\,Palma}
\label{layout}
\vspace*{-.5cm}
\end{figure}

\vspace*{-0.4cm}
The HEGRA (High Energy Gamma Ray Astronomy)\footnotetext{supp.\,by
the BMBF, DFG, CICYT}
 \cite{rev}
experiment is located
at the astronomical observatory of the Canary island La\,Palma
at a height of 2200\,m. It is operated
by a collaboration of the German Max-Planck-Institutes 
f\"ur Kernphysik (Heidelberg)
and f\"ur Physik (M\"unchen),
the univers.\ of Hamburg, Kiel, 
Wuppertal (Germany) and Madrid
(Spain) and the Yerevan Physics Institute (Armenia). The
experimental setup covers a square area of 40000\,m$^2$ 
(Figure\,\ref{layout}). 

\noindent The HEGRA experiment measures extensive air showers 
(EAS) initiated 
by high energetic cosmic photons and nuclei.
It can roughly be divided into two parts:
imaging air Cherenkov telescopes (IACTs) observe selected targets
with an energy threshold of 500 GeV,
while three arrays of detectors monitor the whole sky above the
installation in the energy range beyond 15\,TeV.
\vspace*{-0.5cm} 
\subsection{The HEGRA IACTs}
\vspace*{-0.2cm}
Since the end of 1996 six IACTs are operational.
Four identical telescopes and
an upgraded prototype 
can operate as a system. 
This unique installation observes the
air Cherenkov light images of EAS in a
stereo mode. 
The segmented mirror area of each system telescope is
8.5\,m$^2$ large.
Each camera consist of 271 pixels
(photomultipliers each viewing
0.25$^\circ$ of the sky) covering a field of view of 4.3$^\circ$. 
For photons above the energy threshold of 500\,GeV 
the direction of the primary photon is deduced from the orientation 
of the air Cherenkov light
image in the camera, where the stereo observations 
allow for an angular resolution of better than 0.1$^\circ$.  
Photon induced EAS are separated from showers triggered 
by primary nuclei by the analysis of the image shapes. 
The excellent angular resolution and the efficient photon separation
lead to a nearly background free observation of strong TeV point sources
like the Crab nebula (the galactic standard candle of TeV
astrophysics).
The energy of primary photons can be determined with a resolution
of better than 20\%, thus detailed
energy spectra can be measured \cite{hof}.
The sensitivity of the telescope system can be summarized such,
that within 100\,h observation 
a detection on the 5\,${\rm \sigma}$ level
is possible  for a point source 
with a flux of only 3\% of the Crab.
\vspace*{-0.3cm} 
\subsection{The HEGRA Arrays}
\vspace*{-0.2cm}
The HEGRA arrays sample the EAS showerfront.
243 scintillator huts on a grid with 15\,m spacing 
and a more dense part
around the center register charged particles.
The so called AIROBICC array of 77 stations
(consisting of 20\,cm diameter open photomultipliers attached to
a Winston cone) 
provides a non imaging measurement of the 
air Cherenkov light. 
A Geiger
tower array of 17 stations is located
in the central part of the array.
They allow for the measurement of the electromagnetic energy of an
air shower at detector level and the identification of muons.
The energy threshold 
for vertical photon induced showers was around 20 TeV
for scintillator triggered events, 
which has been lowered to 5\,TeV after a DAQ upgrade in 
spring 1997.
Air showers above 11\,TeV trigger AIROBICC. 
The angular resolution for photon
induced showers at threshold amounts to 0.3$^\circ$ for AIROBICC
triggered events and 0.9$^\circ$ for scintillator data 
(above 20 TeV).
\vspace*{-0.5cm} 
\section{TeV Photon Sources}
\begin{figure}
\vspace*{0.15cm}
\begin{picture}(0,0)(0,0)
\put(1,-0.5){HEGRA}
\end{picture}
\epsfig{file=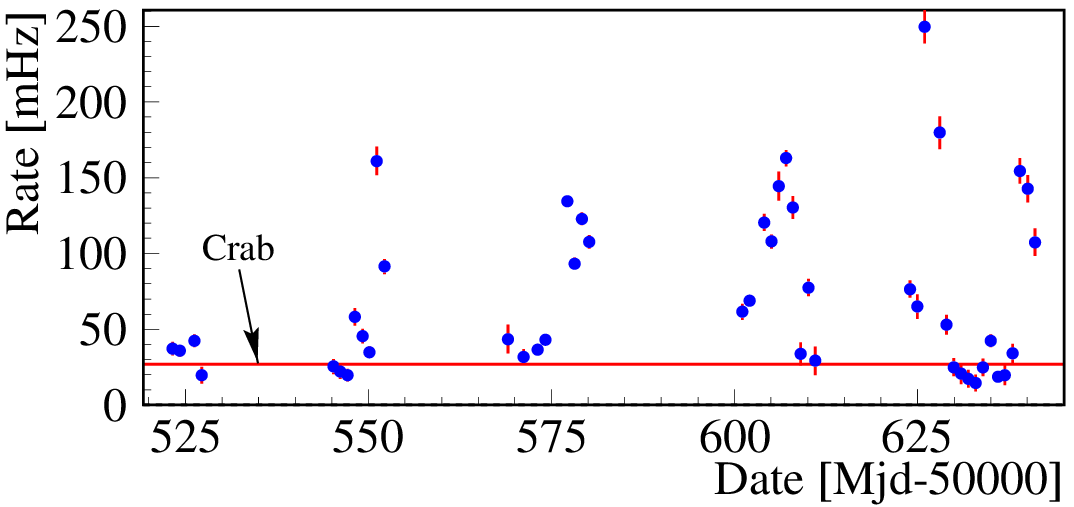,width=5.6cm,angle=0}
\vspace*{-0.6cm}
\caption{\small \em The HEGRA IACT system trigger rate of photons from
Mkn\,501 in a part of 1997. April\,1${\rm ^{st}}$ corresponds to 
539 on the horizontal axis, July\,1${\rm ^{st}}$ to 630.}
\label{flux501}
\vspace*{-0.0cm}
\begin{picture}(0,0)(0,0)
\put(1.7,-0.5){HEGRA}
\end{picture}
\epsfig{file=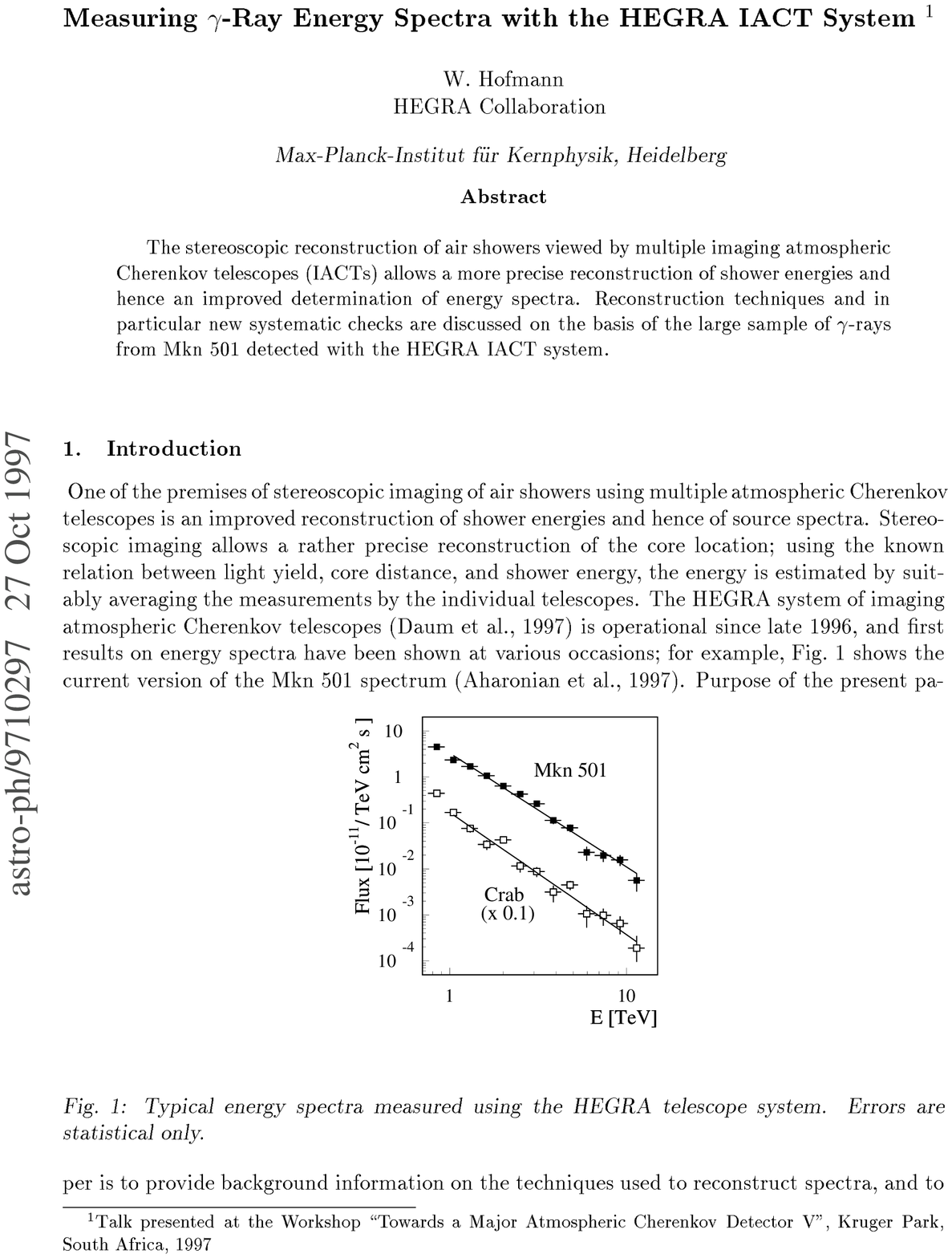,bbllx=210, bblly=210,
bburx=390, bbury=390, clip=, width=5.6cm,angle=0}
\vspace*{-0.6cm}
\caption{\small \em The energy spectra of Mkn\,501 (averaged 
over all measurements in 1997) and of the Crab nebula.
Data beyond 10\,TeV are not shown.}
\label{spec501}
\vspace*{-0.57cm}
\end{figure}

\vspace*{-0.4cm}
The most exciting result in 1997 were the measurements 
with the imaging air Cherenkov telescopes (IACT) of the emission
of TeV photons with a very high and variable flux 
from the extragalactic source Mkn\,501 \cite{501,prot}.
This object in a distance of 500 million lightyears 
belongs to the Blazar class of active galactic nuclei (AGN)
and is thought to contain probably 
a black hole of ${\rm 10^8}$ solar masses.
Blazars are AGNs which emit relativistic jets 
pointed towards the observer.
Under certain conditions 
they can also be detected in the TeV energy
regime.
TeV ${\rm \gamma}$'s from Mkn 501 were first measured in 1995 \cite{whip501}
with a flux corresponding to 8\% of the galactic Crab nebula 
(the strongest known TeV emitter before).
The activity rose to 0.3 Crab in 1996 \cite{brad}.
In 1997 the mean flux lies around 2.0 Crab, but flares with intensities 
of up to 10 times the Crab flux have been registered
approx.\ once a month
(Fig.\,\ref{flux501}).
The energy spectrum in comparison to the spectrum of the 
Crab nebula is shown in Figure\,\ref{spec501}.
The Mkn 501 spectrum follows a power law up to 10\,TeV. 
At higher energies the study of systematic detector effects 
is not finished yet so that the data points are not included 
in the plot.\\
Due to the 
absorption of high energy photons via interaction with 
infrared photons resulting in pair production
these data will provide an important tool to measure indirectly the 
intergalactic infrared background light intensity, which is 
tightly connected to the history of star formation \cite{ir}.
Multiwavelength observations of Mkn 501 \cite{multi}
indicate that the measurements may be explained by 
synchrotron emission of ${\rm e^\pm}$ and inverse 
Compton scattering.
Other sources observed with the Cherenkov telescopes  
are Mkn\,421, an object very similar to Mkn\,501 at nearly
the same distance, but not as spectacular as Mkn\,501 in 1997, 
and the Crab nebula.
No evidence for TeV photon emission could be found from other 
AGNs.
Also no TeV photons could be detected from the supernova remnants
(SNR) IC443 and G78+2. The upper limits provide significant
constraints on models which try to explain the acceleration of the 
bulk of charged cosmic rays in the shells of SNRs.\\
Source searches with the HEGRA arrays 
(with a higher energy 
threshold and less flux sensitivity
compared to the IACTs)
have not been 
successful up to now.
Only small excesses have been found
from the Crab nebula, Hercules\,X-1 and a sample of near Blazars,
but all need further confirmation.
\vspace*{-0.5cm} 
\section{Charged Cosmic Rays} 
\begin{figure}
\vspace*{-1.1cm}
\epsfig{file=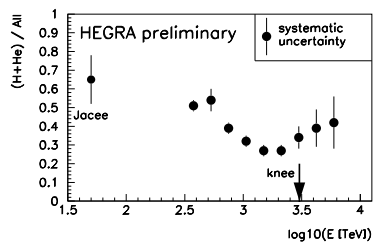,width=5.6cm,angle=0}
\vspace*{-0.6cm}
\caption{\small \em The fraction of light nuclei in cosmic rays.
``JACEE'' marks results obtained by direct balloon measurements.}
\label{compos}
\vspace*{-0.5cm}
\end{figure}

\vspace*{-0.4cm}
The energy spectrum of charged cosmic rays (CR) follows a 
power law 
up to roughly 3\,PeV, where the CR spectrum becomes steeper 
and follows another 
power law. The understanding of 
this so called ``knee'' feature may provide a clue to 
understand the origin of CR.
HEGRA developed new methods to measure the energy    
spectrum and the coarse composition in the energy region around the knee.
One method is described in more detail:
the air Cherenkov light measurements of AIROBICC allow to 
reconstruct the distance between the detector and the air shower maximum.
Together with the registered number of charged particles at detector 
level or the total amount of Cherenkov light the energy contained in the 
electromagnetic shower component can be determined.
This is combined with the energy per nucleon of the primary nucleus 
estimated from the depth of the shower maximum to calculate the 
primary energy and the nucleon number.
While the primary energy can be determined for cosmic ray nuclei 
with an accuracy of 30\%,
the determination of the 
nucleon number is possible only in a coarse manner 
due to natural fluctuations in the
shower developments. Figure\,\ref{compos} shows the  
reconstructed (preliminary) fraction of 
light nuclei (H and He) in the CR.
The results below 1\,PeV are in good agreement to data 
obtained by balloon and satellite measurements. 
The decrease near the knee is in agreement with models describing 
the CR acceleration in supernova remnants, but also a constant
composition is not excluded.
\vspace*{-0.4cm} 
\section{Other HEGRA Results} 
\vspace*{-0.2cm}
No Gamma Ray Burst (GRB) counterparts 
have been found for energies above 5\,TeV,
neither 
coincident with the GRBs or for long time measurements. 
This is in accordance with an extragalactic origin of GRBs,
where no TeV emission is expected due to intergalactic 
absorption of these photons. \\
A small evidence for TeV photons related to the direction 
of the highest energetic (320\,EeV) CR event registered so far \cite{fly}
was detected, which may hint at an intergalactic cascade 
initiated by extremely energetic particles. 
Currently IACT measurements in this region are under way.   
\vspace*{-0.5cm} 
\section{Summary}
\vspace*{-0.3cm}
The last years have established air shower measurements as a new 
branch of high energy astrophysics.
Both TeV ${\rm \gamma}$ observations of extragalactic and galactic
sources as well as analyses of charged cosmic rays 
have been performed successfully and will enhance our understanding
of the nonthermal universe even more in the near future. 
\vspace*{-0.4cm}
%
%

\end{document}